\documentclass[USenglish,twocolumn]{article}

\usepackage[utf8]{inputenc}%(only for the pdftex engine)
\usepackage[big]{dgruyter}
\usepackage{upgreek}

\begin{document}

  \articletype{Research Article{\hfill}Open Access}

\author*[1]{Daria Teplykh}
\author[1]{Valery Malofeev}
\author[1]{Oleg Malov}
\author[1]{Sergey Tyul’bashev}
%affiliation should be placed in lines starting with "{ \let\thempfn\relax%" after \maketitle (see below)
% \affil[1]{Affil, E-mail: email@email.edu}

 %\affil[2]{Affil, E-mail: email@email.edu}

  \title{\huge Peculiarities of radio emission from new pulsars at 111 MHz}

  \runningtitle{Peculiarities of radio emission from new pulsars at 111 MHz}

  %\subtitle{...}

  \begin{abstract}
{The analysis of radio emission of three new pulsars discovered at the Pushchino Radio Astronomy Observatory is presented. The detailed observations were carried out at a frequency of 111 MHz using the Large Phase Array (LPA) and  the standard digital receiver with a total bandwidth is 2.245 MHz and time resolution is 2.46 or 5.12 ms. All pulsars exhibit features of their radiation, the subpulse drift is observed in J0220+3622, the flare activity is exhibited in J0303+2248 and the nulling phenomenon has been detected in J0810+3725.}
\end{abstract}
%  \classification[PACS]{}
 % \communicated{...}
 % \dedication{...}

  \journalname{Open Astronomy}
\DOI{DOI}
  \startpage{1}
  \received{..}
  \revised{..}
  \accepted{..}

  \journalyear{2021}
  \journalvolume{1}
%  \journalissue{1}

\maketitle

%lines connected to author's affiliation

{ \let\thempfn\relax% Remove footnote number printing mechanism
\footnotetext{\hspace{-1ex}{\Authfont\small \textbf{Corresponding Author: Daria Teplykh:}} {\Affilfont P.N. Lebedev Physical Institute of the Russian Academy of Sciences, Leninskii pr. 53, Moscow, 119991, Russia; Email: teplykh@prao.ru}}
}

{ \let\thempfn\relax% Remove footnote number printing mechanism
	\footnotetext{\hspace{-1ex}{\Authfont\small \textbf{Valery Malofeev:}} {\Affilfont P.N. Lebedev Physical Institute of the Russian Academy of Sciences, Leninskii pr. 53, Moscow, 119991, Russia}}
}
{ \let\thempfn\relax% Remove footnote number printing mechanism
	\footnotetext{\hspace{-1ex}{\Authfont\small \textbf{Oleg Malov:}} {\Affilfont P.N. Lebedev Physical Institute of the Russian Academy of Sciences, Leninskii pr. 53, Moscow, 119991, Russia}}
}
{ \let\thempfn\relax% Remove footnote number printing mechanism
\footnotetext{\hspace{-1ex}{\Authfont\small \textbf{Sergey Tyul'bashev:}} {\Affilfont P.N. Lebedev Physical Institute of the Russian Academy of Sciences, Leninskii pr. 53, Moscow, 119991, Russia}}
}

\section{Introduction}

For many years the interest of the neutron stars observers fueled by the discovery of new pulsars in different wavelengths, as well as by a wide variety of observational features of these sources. Radio pulsars are generally known as highly variable objects. Fluctuations in the flux densities vary over a wide time range from fractions of a microsecond to several years. They can be caused by various factors: both external (for example, scintillation in the interstellar medium) and internal (flare activity, nulling, mode switching, etc., (see, for example, the monographs of \cite{PSR}, \cite{LK2004}). Despite a half-century pof pulsar research, the mechanism of pulsar emission remains unknown. Therefore, the study of some peculiarities in their radiation, detected in new pulsars, can help to clarify some fundamental points of the emission mechanism and the structure of the magnetosphere of these sources.

Since 2014, a pulsar search program is carried out by the upgraded LPA (Large Phase Array) radio telescope in Pushchino Radio Astronomy Observatory, thereby the discovery of more than 70 new pulsars and RRATs (Rotating Radio Transients) (\cite{TT15a,TT15b,TTO16,TTK17,T18AR,T18AA,TKT20}) has been made. The Pushchino pulsar search program is based on daily round-the-clock monitoring of a large area of the sky ($-9^{\circ} <\delta <42^{\circ}$). This approach is good for searching for weak sources and the objects with variable radiation, both on short time scales (milliseconds - minutes) and long ones (several days - years). A detailed study of radio emission from new sources revealed a number of features in their emission such as the nulling phenomenon, flare activity, and drift of subpulses. In this paper, we present observations and brief analysis of the above-mentioned features of the radio emission from three new Pushchino pulsars.

\section{Observations}\label{sect:Obs}

The observations were carried out in the Pushchino Radio Astronomy Observatory at a frequency of 111 MHz using the meridian radio telescope LPA. Its antenna is the phased array composed of 16384 dipoles. The geometric area of this antenna is about  $70 000m^{2}$ and the effective area is   $47000 \pm 2500\,m^{2}$ (\cite{TTO16}). Antenna has 128 space beams with the size of one beam $0.5^{\circ} \times 1^{\circ}$. The duration of observing session is $3.5/cos\,\delta$. During  the search, the data is recorded simultaneously in two modes: 6 channels with bandwidth of 400 kHz each, with the sample of 100 ms; or 32 channels (75 kHz each), with a sample of 12.5 ms. For detailed studying of known pulsars, the observations are carried out using the standard digital receiver with a high frequency-time resolution: 470 channels $\times$ 4.88 kHz and the total bandwidth is 2.245 MHz., time resolution is 2.46 or 5.12 ms. All data is stored in the server. For their processing the special program has been worked out (\cite{MTL12}). 

\section{Results}\label{sect:Res}
\subsection{J0220+3622, pulsar with the subpulse drift}

Subpulse drift is a sequence of individual pulses/subpulses shifting in phase from one edge of the mean pulse profile to the opposite, forming characteristic drift bands on the longitude-time diagram (\cite{DC68}). This  phenomenon is characterized by two periods: second ($P_{2}$) and  third ($P_{3}$) class. The value of $P_{2}$ is the horizontal drift band separation in time units. $P_{3}$ is a distance which is determined by the number of periods $P$ on the ordinate. The subpulse drift phenomenon has been repeatedly investigated earlier by both observers and theorists. This effect is caused, probably, by the movement of emission regions in the pulsar magnetosphere and is associated with the radiation mechanism. A complete physical understanding of this phenomenon is still lacking, but the most famous explanation is the model of \cite{RS75}, which was subsequently extended by many authors (for example, \cite{FR82}, \cite{DR99}, \cite{GMG03}), explaining the drift phenomenon by rotating the sub-beams around the magnetic axis (“carousel” model). The model describes the effect well for pulsars with single pulse profiles, since the subpulse drift is believed to be related to conal radiation, and it is much more complicated when the drift phenomenon is present in pulsars with complex, multicomponent pulse profiles (\cite{R86}).

\begin{figure}
	\framebox[1.02\columnwidth]{\Huge 	\includegraphics[width=1\columnwidth]{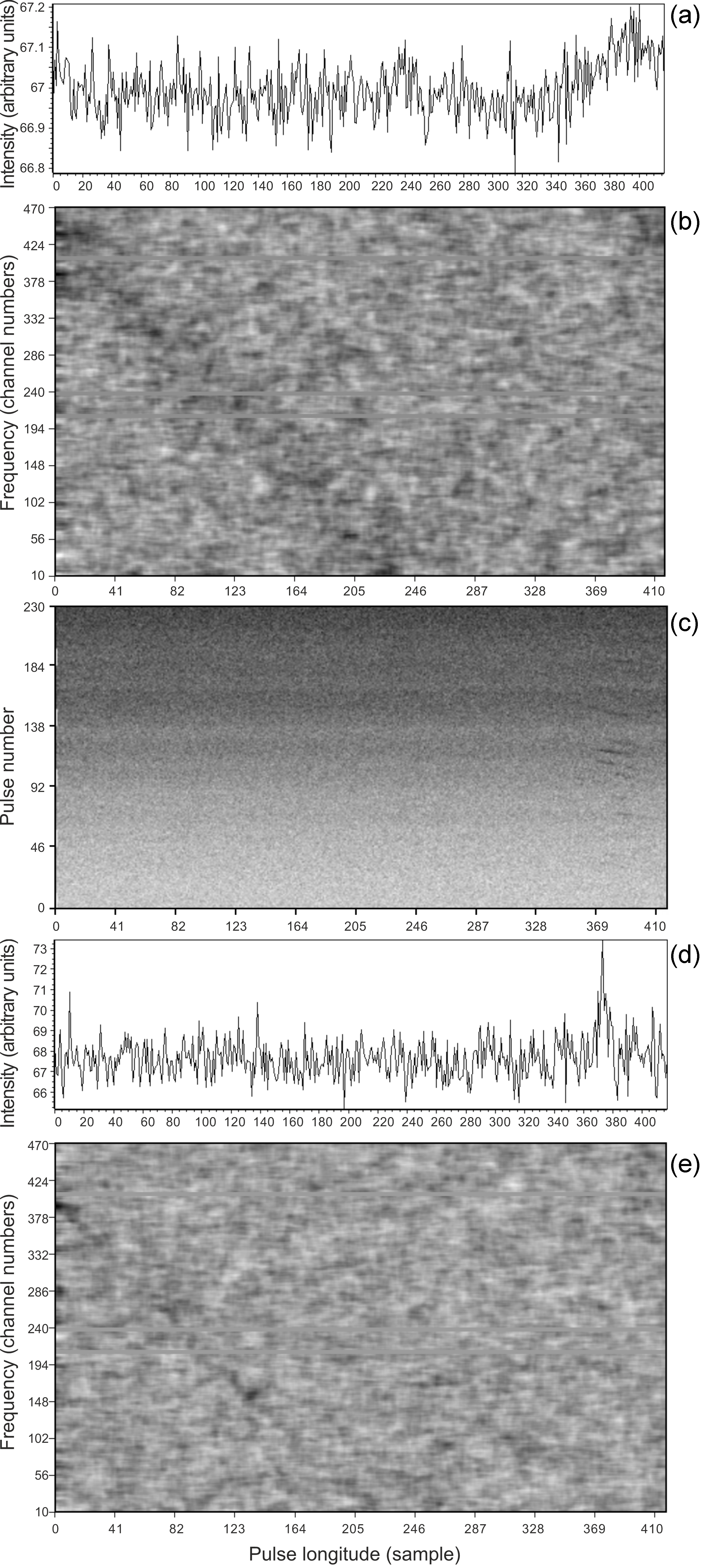}}
	
	\caption{The  example  of  observations  for  pulsar  J0220+36  02.01.2018.  From  top  to bottom: a) integrated pulse profile, b) dynamic spectrum and c) variations of pulse intensity during one observation session (230 pulses), d) example of individual pulse profile, e) dynamic spectrum for individual pulse. The abscissa axis for all graphs shows pulse period (one sample is equal to 2.46ms).
	\label{fig:fig1}}
\end{figure}

\begin{figure}
	\framebox[1.02\columnwidth]{\Huge 	\includegraphics[width=1\columnwidth]{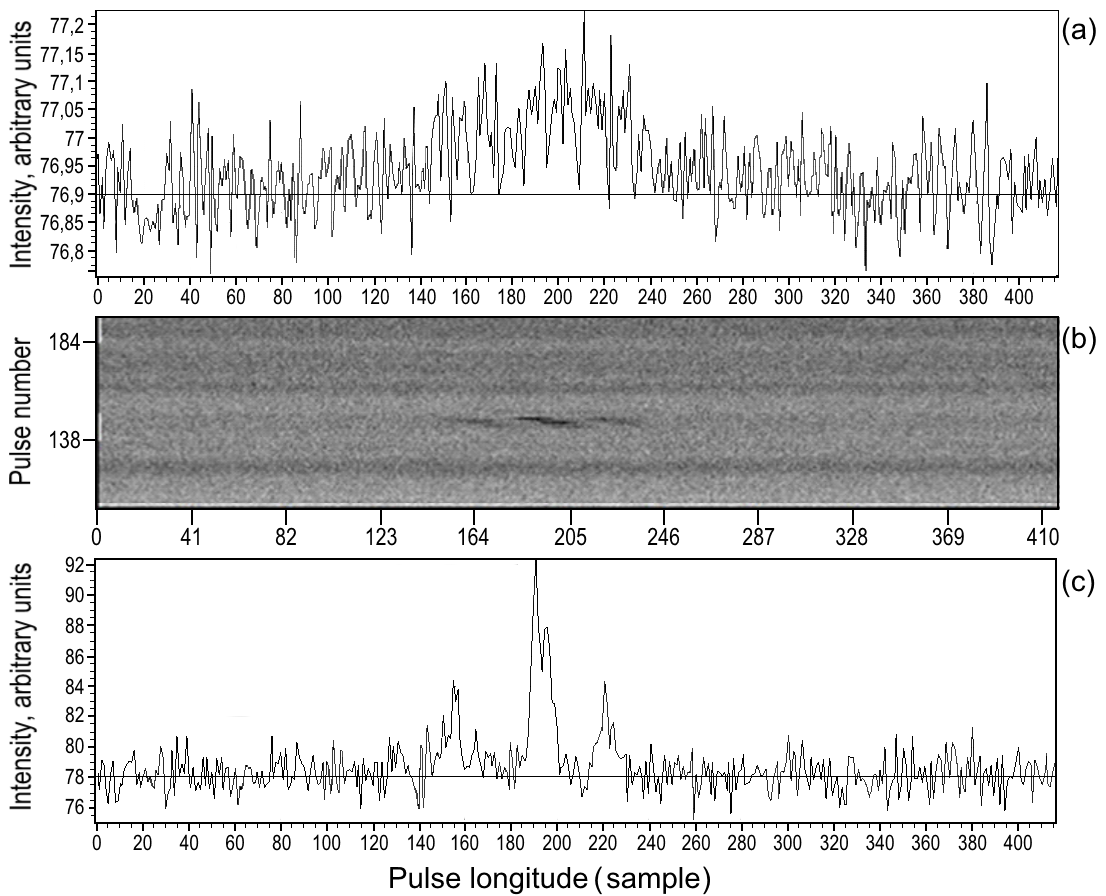}}
	
	\caption{a)  – the integrated  profile  of  the  observed  pulsar;  b)  – the  pulse intensity  dependence for the 100 periods, drift of the strong pulses is visible; c) – pulse profile of one of three strong individual pulses).
	\label{fig:fig2}}
\end{figure}

\begin{figure}
	\framebox[\columnwidth]{\Huge 	\includegraphics[width=\columnwidth]{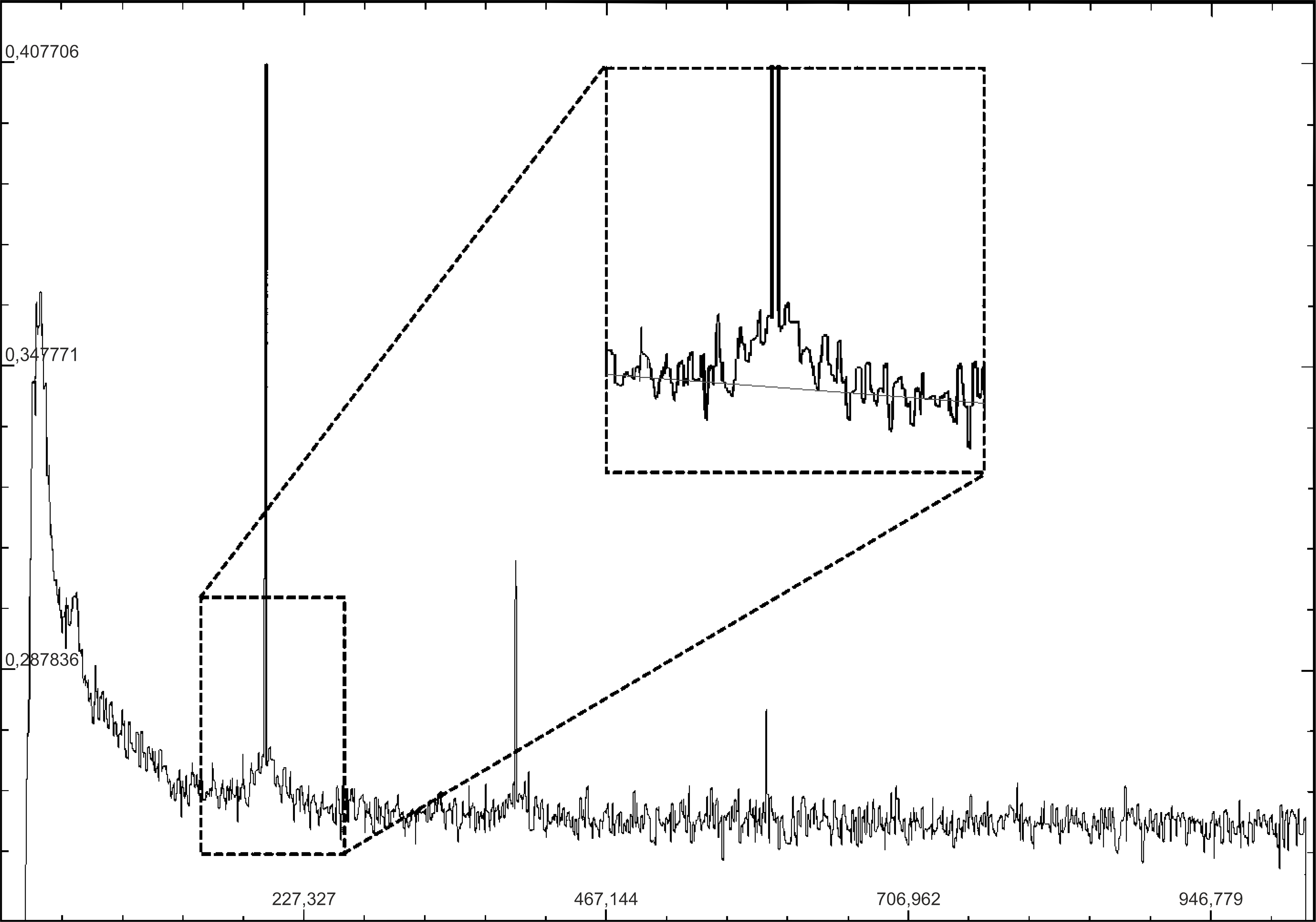}}
	
	\caption{The summed Fourier spectrum of J0220+36 obtained by accumulation more than 500 daily Fourier spectra.
	\label{fig:fig3}}
\end{figure}

\begin{figure}
	\framebox[\columnwidth]{\Huge \includegraphics[width=\columnwidth]{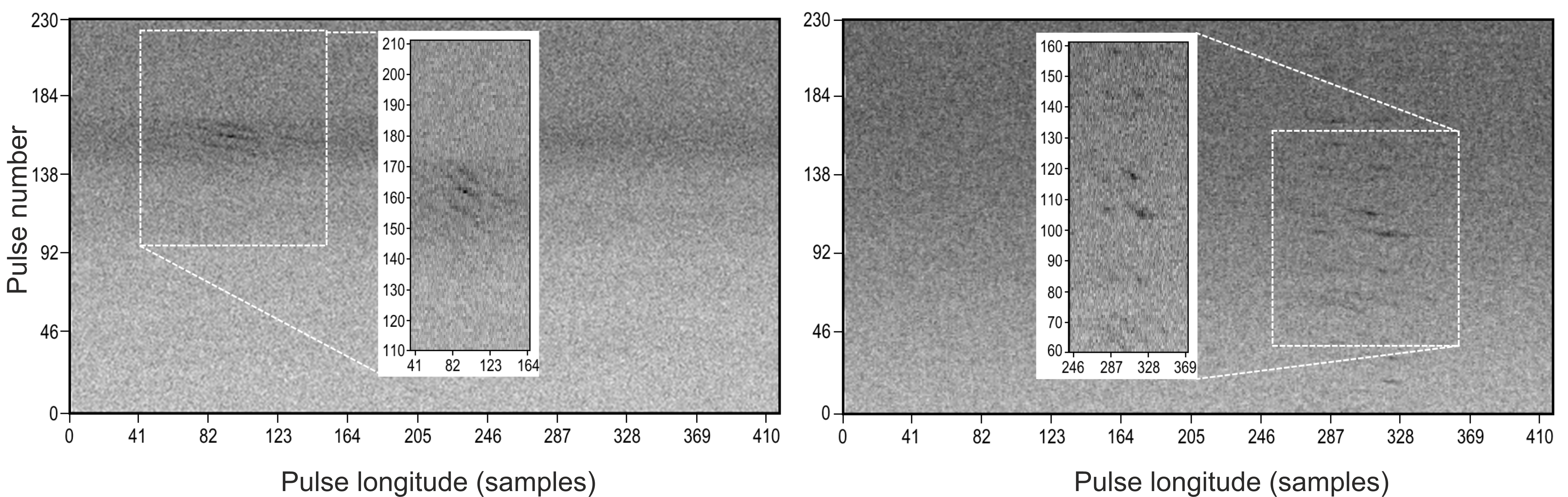}}
	\caption{The  examples  of the drift  for  J0220+36  on  different  observation  days.
	\label{fig:fig4}}
\end{figure}

The pulsar J0220+36 with the period $P = 1.0297 s$ and the dispersion measure $DM = 46 \pm 1\,pc/cm^{3}$ (\cite{TT15a}) has a very wide average pulse profile of $\sim$220 ms and narrow individual pulses (Fig.\ref{fig:fig1}). Pulsar has a complex multicomponent structure of individual pulses, drift of subpulses, as well as short bursts of radiation, when the drift becomes especially clear (\cite{TMM20}). During one of 222 observation days (November 3, 2017), the pulsar showed the increasing of the emission activity (Fig.\ref{fig:fig2}). A three-component pulse structure (Fig.\ref{fig:fig2}c) is visible in the series of three consecutive pulses with the S/N from 8 to 16 (Fig.\ref{fig:fig2}b). The distances between subpulses are on average 82 ms and 64 ms. The analysis of time intervals between subpulses of individual pulses made it possible to measure the periods of the second and the third classes: $P_{2} = 70\pm 10$ ms, $P_{3} = 7\pm1$, as well as the drift rate $D = 20\pm5$ ms per period. There is new method to measure periods $P_{2}$ and $P_{3}$ (\cite{MT18}). Fig.\ref{fig:fig3} shows the summed power spectrum for J0220+3622 obtained by adding more than 500 daily Fourier spectra. The presence of a third class period is indicated by a significant substrate (pedestal) at the first two harmonics of the summed Fourier spectrum. As follows from the work of \cite{MT18}, the period of the third class can be calculated from the position of the satellites near the harmonics. But since the summed power spectrum of this object does not show evident satellites, we can estimate only the low limit of $P_{3}$ ($P_{3}>6$). This pulsar exhibits a rare phenomenon:  a variable drift velocity, which has been first noted for PSR B0031-07 (\cite{HTT70}). Figure \ref{fig:fig4} demonstrates of the subpulse drift for two days of observations.

\subsection{J0303 + 2248, pulsar with flare activity}

The pulsar J0303 + 2248 (\cite{TT15a}) discovered with $P=1.207$ and $DM=19$ has irregular, flare activity. This type of sporadic signal amplification is observed in a number of radio pulsars (for example, as giant pulses (\cite{SSP71}), and as the characteristic of radio transients (\cite{M2006}).

\begin{figure}
	\framebox[\columnwidth]{\Huge 	\includegraphics[width=\columnwidth]{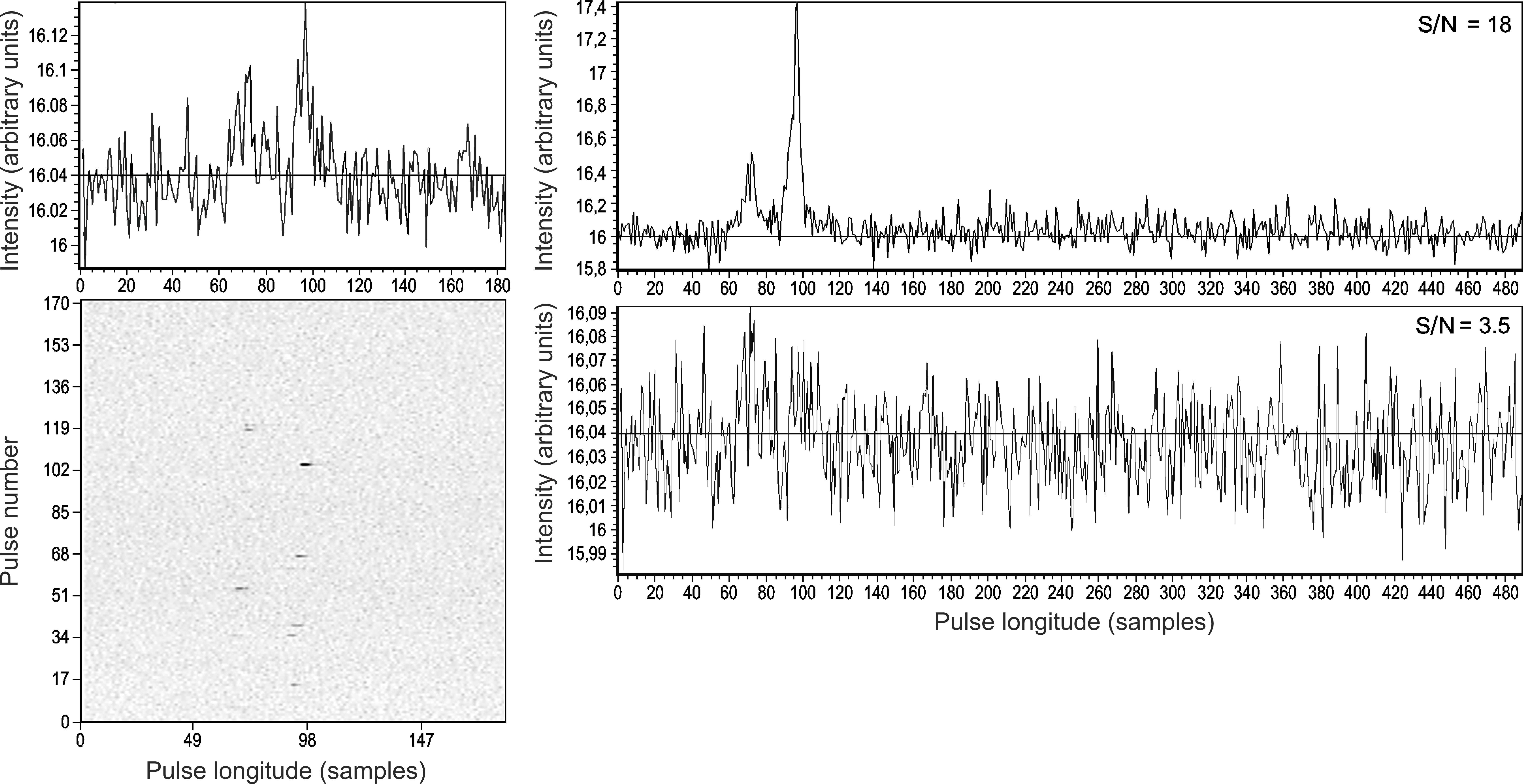}}
	\caption{The example of observations of J0303+22 (25.01.2019). The integrated pulse profile (top left) and intensity changes of the pulsar pulses during one observation session (bottom left). The pulse profiles of a pulsar obtained by summing 8 strong pulses with $S/N>5$ (top right) and by summing of all other pulses with $S/N\leq5$ (bottom right).
	\label{fig:fig5}}
\end{figure}

\begin{figure}
	\framebox[\columnwidth]{\Huge 	\includegraphics[width=\columnwidth]{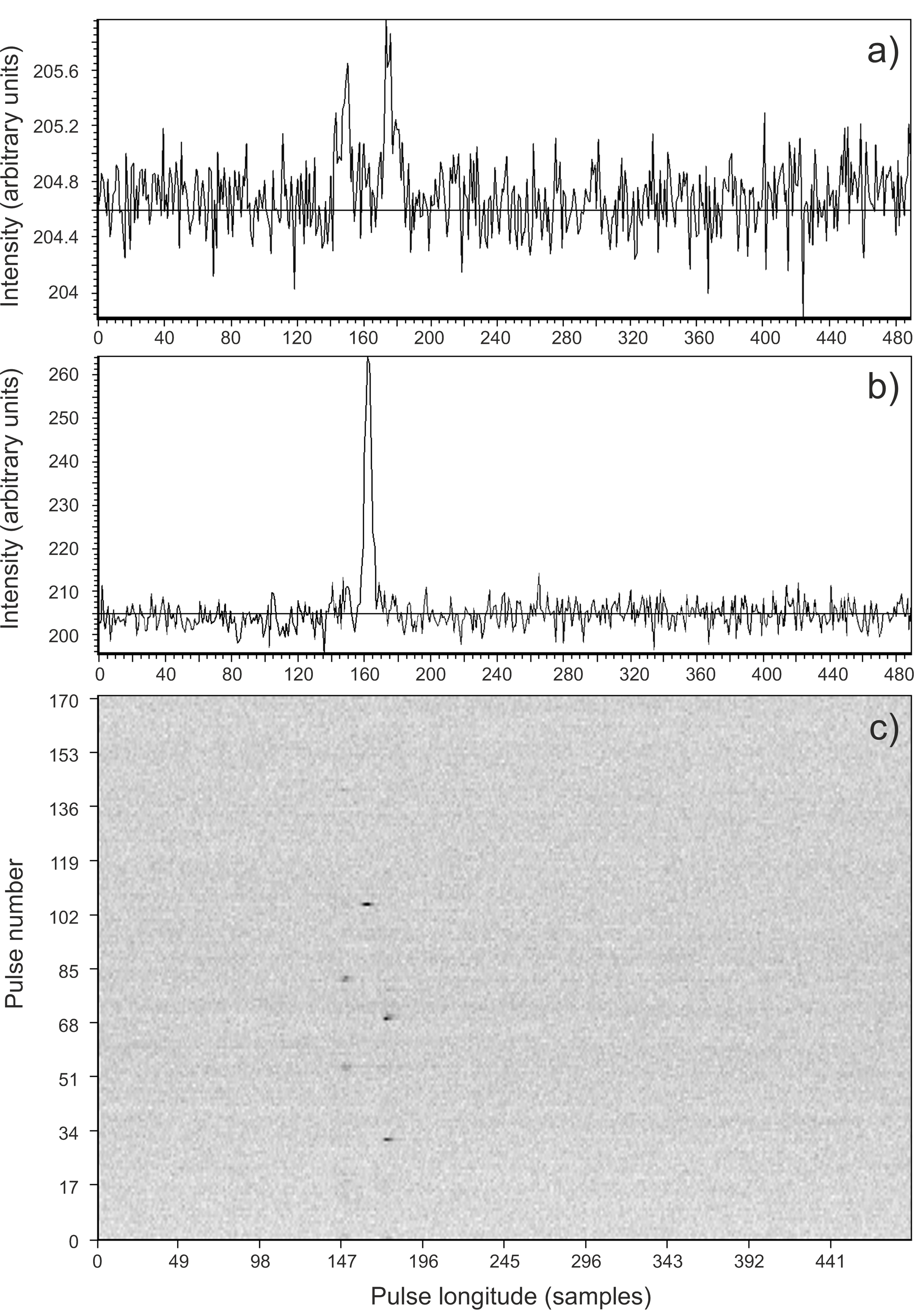}}
	\caption{The inter-component emission detection from J0330+2248 (13 January 2018. (a) The integral pulse profile, (b) profile of the individual pulse (number 105) with S/N = 20.5 corresponding to inter-component emission, (c)  the intensity changes of the pulses during the observation session.
	\label{fig:fig7}}
\end{figure}

A more detailed study showed that the pulsar J0303+2248 exhibits powerful single pulses, and its emission is more similar to radiation from transient sources (RRATs), but with more frequent pulses.  This pulsar has a two-component pulse structure with rare and weak inter-component emission. There is practically no radiation outside strong pulses. Fig.\ref{fig:fig5} shows the sum of the eight strongest individual pulses. The signal-to-noise ratio (S/N) of this summed pulse is five times more than the S/N for sum of other 160 pulses. Some individual pulses are ten times stronger than the integrated pulse. The annual variations of the radio emission show that the signal has been recorded at the level S/N = 3-7 in most of the observation sessions. But we can see the increasing of the signal on some days. Also, this object shows a very rare phenomenon (one event during 146 days of observations). It is a flare in the inter-component space (Fig.\ref{fig:fig7}). A similar behavior was found earlier for the pulsar J0653 + 8041 (\cite{MT16}), where the central component flares were observed very rarely in the three-component profile. We obtained more precise value of the dispersion measure ($DM$) for J0303+2248,  it was $19\pm1 pc/cm^{3}$.

\onecolumn

\twocolumn

\subsection{J0810 + 3725 - nulling pulsar}

The nulling is a phenomenon when we can observe a temporary absence of pulsed emission from a neutron star (\cite{Backer1970}). Long-term studies of this phenomenon have shown that the percentage of nulling fraction (NF) can vary in range 1\% - 90\% (\cite{Wang2007,Gajjar2014,Burgay2019}). Intermittent pulsars belong to the separate class of objects switching off their emission for a longer time (from several days to several years) (\cite{Kramer2006,Camilo2012}). 

\begin{figure}
	\framebox[\columnwidth]{\Huge 	\includegraphics[width=\columnwidth]{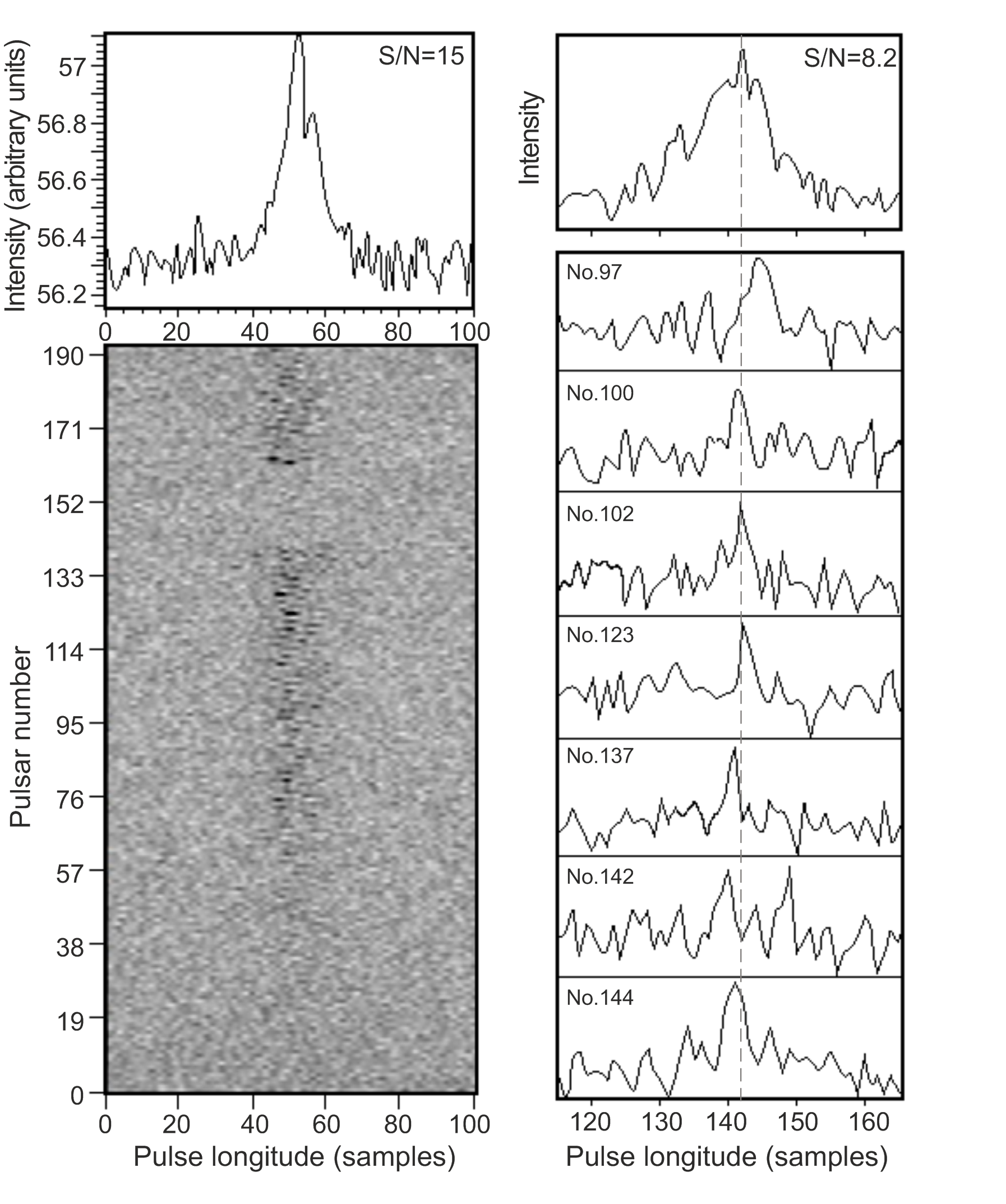}}
	\caption{The example of the pulse profile of J0810 + 37 and the intensity of individual pulses during one observation session (left); and examples of individual pulse profiles (right).
	\label{fig:fig8}}
\end{figure}

The preliminary results of the J0810 + 3725 study were presented in the work (\cite{TM2019}). The pulsar with $P=1.2482$ and $DM$= 17 (\cite{TT15a})  has a wide emission window (the average half-width of the integral pulse is $W_{0.5} \sim$ 25 ms), while the individual pulses are very narrow ($W_{0.5} \sim$ 8 ms) (Fig. \ref{fig:fig8}).

\begin{figure}
	\framebox[\columnwidth]{\Huge 	\includegraphics[width=\columnwidth]{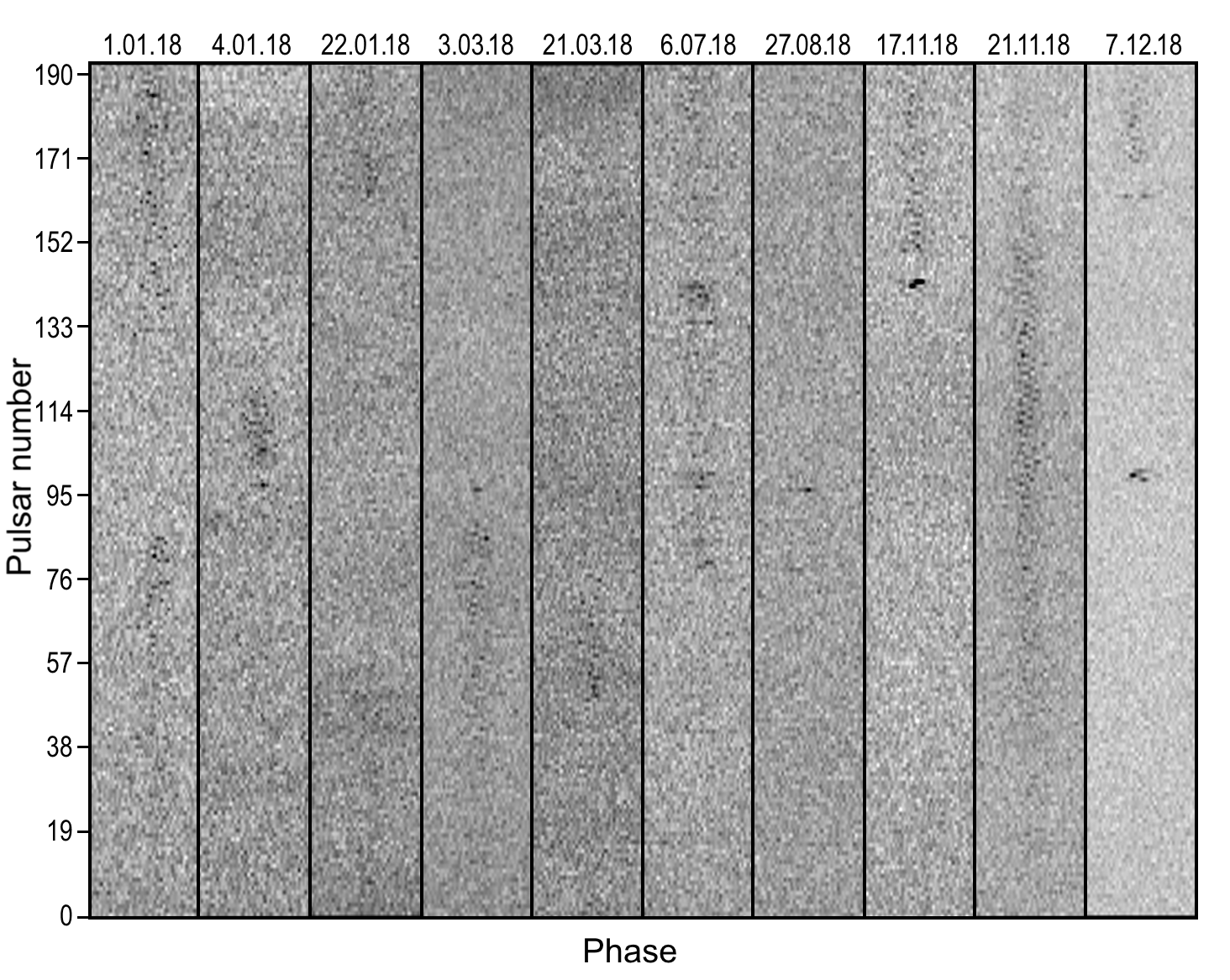}}
	\caption{The examples of nulling for the pulsar J0810+37 on different observation days.
	\label{fig:fig9}}
\end{figure}

Examples of nulling for the pulsar J0810 + 3725 are shown on Fig. \ref{fig:fig9}. In active state (during periods of “switch on”) the pulsar demonstrates a wide spread of nulling durations (10\% - 90\% of the observation session time) and in average is 40\% (\cite{TM2019}). During some days (Fig. \ref{fig:fig8}, \ref{fig:fig9}) we can see the allusion, that this pulsar has a subpulse drift. We shall try estimate $P_{2}$ and $P_{3}$ in the future. Pulsar manifests intermediate object between two classes: ordinary nullers (eg, \cite{Wang2007}) and “switch off” pulsars (\cite{Kramer2006,Camilo2012}). We can make this conclusion on the base of our investigation. This pulsar demonstrates the emission interruption for various intervals of the time: from several pulses to several days.

\section{Conclusion}

The detailed analysis of the study of radio emission from new pulsars is carried out. As expected, the Pushchino search program is sensitive to sources with peculiar emissions. Some of the new objects have interesting radio emission features such as nulling, subpulse drift, and flare activity. 

Pulsar J0220+3622 demonstrates the subpulse drift with a variable drift rate. Pulsar J0303+2248 has the flare activity and very rare case of the flare in the inter-component space.  

Pulsar J0810+3725 shows the intermediate nulling phenomenon with the visible  subpulse drift.


\begin{thebibliography}{30}
    \bibitem[Backer, 1970]{Backer1970}    Backer, D. C., \textbf{1970}, \textit{Nature}, 228, 42.

    \bibitem[Burgay et al., 2003]{Burgay2019}  Burgay, M., Stappers, B., Bailes, M. et al., \textbf{2019}, \textit{MNRAS}, 484, 5791.
    
    \bibitem[Deshpande \& Rankin, 1999]{DR99} Deshpande, A.~A. \& Rankin, J.~M.\textbf{1999}, \textit{ApJ}, 524, 1008.
    
    \bibitem[Drake \& Craft, 1968]{DC68} Drake, F.~D. \& Craft, H.~D. \textbf{1968}, \textit{Nature}, 220, 231. 
    
    \bibitem[Gajjar et al., 2014]{Gajjar2014} Gajjar, V.,  Joshi, B. C., Wright, G. et al.  \textbf{2014}, \textit{MNRAS}, 439, 221.
    
    \bibitem[Camilo et al., 2012]{Camilo2012} Camilo, F., Ransom, S. M., Chatterjee, S. et al., \textbf{2012}, \textit{ApJ}, 746, 63.
    
    \bibitem[Filippenko \& Radhakrishnan, 1982]{FR82} Filippenko, A.~V. \& Radhakrishnan, V. \textbf{1982}, \textit{ApJ}, 263, 828.
    
    \bibitem[Gil et al., 2003]{GMG03} Gil, J., Melikidze, G.~I., \& Geppert, U.\textbf{2003}, \textit{A\&A}, 407, 315.
    
    \bibitem[Huguenin et al., 1970]{HTT70} Huguenin, G.~R., Taylor, J.~H., \& Troland, T.~H. \textbf{1970}, \textit{ApJ}, 162, 727.
    
    \bibitem[Kramer et al., 1970]{Kramer2006}
    Kramer, M., Lyne, A. G., O'Brien, J. T. et al., \textbf{2006}, \textit{Science}, 312, 549

	\bibitem[Lorimer \& Kramer, 2004]{LK2004} Lorimer, D.~R. \& Kramer, M.\ \textbf{2004}, Handbook of pulsar astronomy, Cambridge observing handbooks for research astronomers,  \textit{Cambridge University Press}, UK, Cambridge.

	\bibitem[Malofeev et~al., 2012]{MTL12} 	Malofeev, V.~M., Teplykh, D.~A., \& Logvinenko, S.~V. \textbf{2012}, \textit{Astron. Rep.}, 56, 35.
	
	\bibitem[Malofeev et al., 2016]{MT16} Malofeev, V.~M., Teplykh, D.~A., Malov, O.~I., et al. \textbf{2016}, \textit{MNRAS}, 457, 538.
	
	\bibitem[Malofeev \& Tyul'bashev, 2018]{MT18} Malofeev, V.~M. \& Tyul'bashev, S.~A.\textbf{2018}, \textit{RAA}, 18, 096.

	\bibitem[Manchester \& Taylor(1977)]{PSR} Manchester, R.~N. \& Taylor, J.~H., Pulsars, \textbf{1977},  \textit{W. H. Freeman}, San Francisco.	

	\bibitem[McLaughlin et al., 2006]{M2006} McLaughlin, M.~A., Lyne, A.~G., Lorimer, D.~R., et al.\textbf{2006}, \textit{Nature}, 439, 817.
	
	\bibitem[Rankin, 1986]{R86} Rankin, J.~M. \textbf{1986}, \textit{ApJ}, 301, 901.
	
	\bibitem[Ruderman \& Sutherland, 1975]{RS75} Ruderman, M.~A. \& Sutherland, P.~G. \textbf{1975}, \textit{ApJ}, 196, 51. 
	
	\bibitem[Sutton et al., 1971]{SSP71} Sutton, J.~M., Staelin, D.~H., \& Price, R.~M. \textbf{1971}, The Crab Nebula, \textit{Proceedings from IAU Symposium}, 46, 97.
	
	\bibitem[Teplykh \& Malofeev, 2019]{TM2019} Teplykh, D. A., Malofeev, V. M. \textbf{2019}, \textit{Bull. Lebedev Phys. Inst.}, 46, 380.
	
	\bibitem[Teplykh et al., 2020]{TMM20} Teplykh, D., Malofeev, V., \& Malov, O. \textbf{2020}, \textit{Ground-Based Astronomy in Russia. 21st Century}, 446.
	
	\bibitem[Tyul'bashev \& Tyul'bashev, 2015a]{TT15a} Tyul'bashev, S.~A. \& Tyul'bashev, V.~S. \textbf{2015}, \textit{Astronomicheskij Tsirkulyar}, 1624.
	
	\bibitem[Tyul'bashev \& Tyul'bashev, 2015b]{TT15b} Tyul'bashev, S.~A. \& Tyul'bashev, V.~S. \textbf{2015}, \textit{Astronomicheskij Tsirkulyar}, 1625.
	
	\bibitem[Tyul'bashev et al., 2016]{TTO16} Tyul'bashev, S.~A., Tyul'bashev, V.~S., Oreshko, V.~V., \& Logvinenko, S.~V.	\textbf{2016}, \textit{Astron. Rep.}, 60, 220.
	
    \bibitem[Tyul'bashev et al., 2017]{TTK17} Tyul'bashev, S.~A., Tyul'bashev, V.~S., Kitaeva, M.~A., et al. \textbf{2017}, \textit{Astron. Rep.}, 61, 848. 	
    
    \bibitem[Tyul'bashev et al., 2018a]{T18AR} Tyul'bashev, S.~A., Tyul'bashev, V.~S., Malofeev, V.~M., et al. \textbf{2018}, \textit{Astron. Rep.}, 62, 63. 
    
    \bibitem[Tyul'bashev et al., 2018b]{T18AA} Tyul'bashev, S.~A., Tyul'bashev, V.~S., \& Malofeev, V.~M. \textbf{2018}, \textit{A\&A}, 618, A70. 
    
	\bibitem[Tyul'bashev et al., 2020]{TKT20} Tyul'bashev, S.~A., Kitaeva, M.~A., Tyul'bashev, V.~S., et al. \textbf{2020}, \textit{Astron. Rep.}, 64, 526.
	
	\bibitem[Wang et al., 2007]{Wang2007}  Wang, N., Manchester, R. N., \& Johnston, S. \textbf{2007}, \textit{MNRAS}, 377, 1383.
	



\end{thebibliography}
\end{document}